\documentclass[letter]{aa}
\usepackage{graphicx}
\usepackage{amsmath, bm}
\usepackage{natbib}
\bibpunct{(}{)}{;}{a}{}{,} 

\usepackage{txfonts}
\usepackage{longtable}
\usepackage{listings}
\usepackage{color}
\usepackage{textcomp}

\defcitealias{Cantat-Gaudin2020b}{CGa20}

\begin{document}

\title{The star cluster age function in the Galactic disc with {\it Gaia} DR2}
\subtitle{Fewer old clusters and a low cluster formation efficiency}

\author{Friedrich Anders\inst{1}, Tristan Cantat-Gaudin\inst{1}, Irene Quadrino-Lodoso\inst{1},\\ Mark Gieles\inst{1,2}, Carme Jordi\inst{1}, Alfred Castro-Ginard\inst{1}, Lola Balaguer-Núñez\inst{1}
        }
    
\institute{Dept. FQA, Institut de Ci\`encies del Cosmos, Universitat de Barcelona (IEEC-UB), Mart\'i i Franqu\`es 1, E-08028 Barcelona, Spain\\
\email{\{fanders;tcantat\}@fqa.ub.edu}
\and{ICREA, Passeig Lluis Companys 23, E-08010 Barcelona, Spain}
           }

\date{Received May 29, 2020; accepted November 27, 2020}

  \abstract{We perform a systematic reanalysis of the age distribution of Galactic open star clusters. Using a catalogue of homogeneously determined ages for 834 open clusters contained in a 2 kpc cylinder around the Sun and characterised with astrometric and photometric data from the {\it Gaia} satellite, we find that it is necessary to revise earlier works that relied on data from the Milky Way Star Cluster survey. After establishing age-dependent completeness limits for our sample, we find that the cluster age function in the range $6.5 < \log t<10$ is compatible with Schechter-type or broken power-law functions. 
Our best-fit values indicate an earlier drop of the age function (by a factor of $2-3$) with respect to the results obtained in the last five years, and are instead more compatible with results obtained in the early 2000s along with radio observations of inner-disc clusters. Furthermore, we find a typical destruction timescale of $\sim1.5$ Gyr for a $10^4\, {\rm M}_{\odot}$ cluster and a present-day cluster formation rate of $0.55_{-0.15}^{+0.19}$ Myr$^{-1}$kpc$^{-2}$, suggesting that only $16_{-8}^{+11}$ \% of all stars born in the solar neighbourhood form in bound clusters. 
Accurate cluster-mass measurements are now needed to place more precise constraints on open-cluster formation and evolution models.}

\titlerunning{The cluster age function of the Milky Way}
\authorrunning{Anders, Cantat, Quadrino, et al.}
\maketitle

\section{Introduction}

It is becoming increasingly difficult to understand the formation of galaxies without taking into account several levels of baryonic hierarchical structure formation. To unravel the formation history of the Milky Way disc, however, it is often useful to study open star clusters (OCs): groups of stars of the same age and abundance pattern held together by mutual gravitation.

\begin{figure}
\includegraphics[trim=0 145 0 170, clip, width=.46\textwidth]{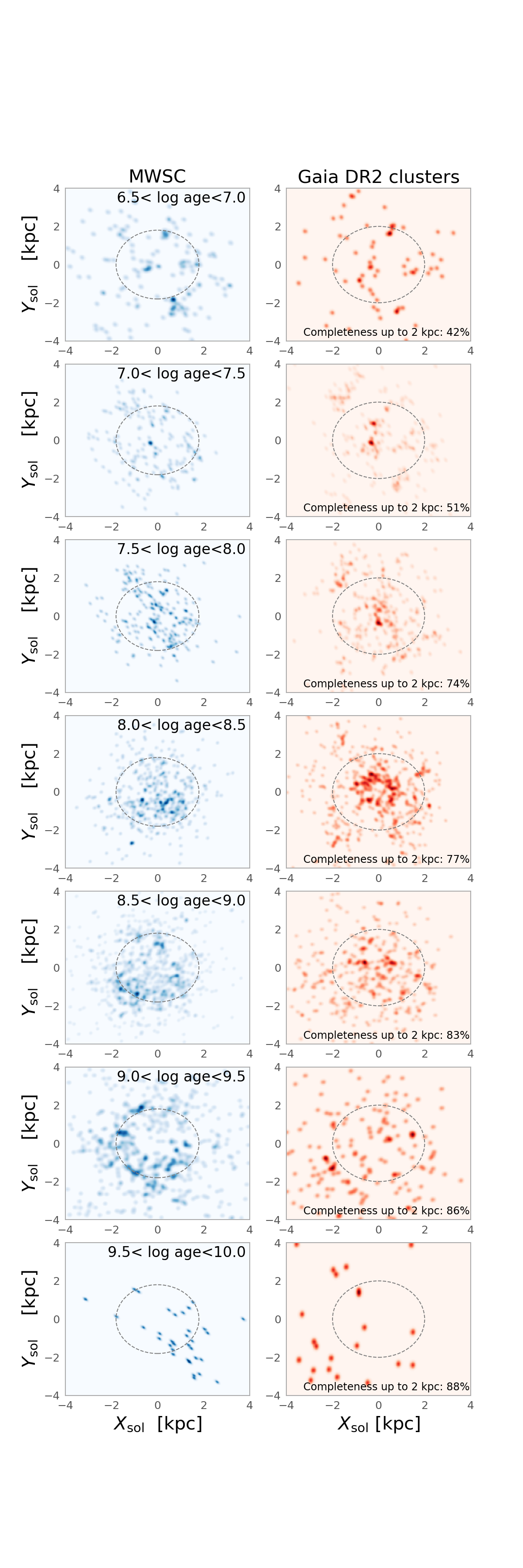} 
\caption{\label{fig:xymap} 
Galactic distribution of the OC samples studied in this Letter, sliced into logarithmic age bins. Left: Pre-{\it Gaia} census using the MWSC catalogue \citep{Kharchenko2013}. Right: Post-{\it Gaia} DR2 census, using the catalogue of \citet{Cantat-Gaudin2020b}. In each panel we show a 2D kernel density estimate with a fixed bandwidth of 0.05 kpc. For the MWSC, the dashed circle corresponds to the completeness limit of 1.8 kpc used in the literature (e.g. \citealt{Piskunov2018}); for the {\it Gaia} DR2 census, a sample limit of 2 kpc is used together with age-dependent completeness fractions indicated in each panel (see Sect. \ref{sec:data}, last paragraph).} 
\end{figure}

The physical processes governing the formation and evolution of OCs are encoded in the distribution of their properties, including mass, age, and size (for a recent review, see \citealt{Krumholz2019}). Since it is relatively easy to estimate at least differential ages for OCs, one of the key observables of the local OC population is the completeness-corrected age distribution \citep[e.g.][]{Wielen1971, Janes1988, Battinelli1991, Lamers2005, Piskunov2006, Morales2013, Piskunov2018, Krumholz2019}. This cluster age function (CAF) can be thought of as an integral of the cluster distribution function over several other parameters, such as present-day mass, initial mass, internal rotation, and binary fraction,  which are much more difficult to determine.

In the Milky Way, the census of OCs is highly incomplete, at least beyond a local volume of $1-2$ kpc \citep{Kharchenko2013}. Thanks to the unprecedented quality of the astrometric and photometric data released with the {\it Gaia} second data release (DR2; \citealt{GaiaCollaboration2018}), hundreds of new clusters have recently been detected even at smaller distances \citep[e.g.][]{Cantat-Gaudin2018, Cantat-Gaudin2019, Castro-Ginard2019, Castro-Ginard2020, Liu2019, Sim2019}. In addition, some analyses have shown that previous catalogues also contained large numbers of false positives and asterisms (\citealt{Cantat-Gaudin2018, Cantat-Gaudin2020a}). The impact of {\it Gaia} on the field of Galactic cluster studies can thus hardly be overestimated.

The main remaining challenges for obtaining a clean and unbiased CAF for the Milky Way (or at least for the local solar neighbourhood of a few kiloparsec) are a) the irregular dust distribution in the Galactic disc; b) the intrinsically patchy distribution of star clusters and other young disc tracers \citep{Becker1963, Becker1970, Efremov2010, Moitinho2010, Cantat-Gaudin2018, Reid2019, Skowron2019}, rendering completeness estimates difficult; c) the smooth transition between moving groups, associations, and physically bound OCs \citep{Krumholz2019, Kounkel2019, Cantat-Gaudin2020a, Kounkel2020}; and d) the availability of homogeneously derived cluster parameters. 

The last problem has recently been addressed by \citet[][hereafter CGa20]{Cantat-Gaudin2020b} who published a catalogue of homogeneous age estimates for 1\,867 Galactic OCs confirmed by {\it Gaia} DR2. In this Letter, we use that catalogue to reevaluate the CAF of the Milky Way CAF. Our figures (including the completeness analysis) are reproducible via the {\tt python} code provided online\footnote{\url{https://github.com/fjaellet/gaidr2-caf}}.

\section{The {\it Gaia} DR2 open-cluster census} \label{sec:data}

The precise {\it Gaia} DR2 astrometry (positions, proper motions, and parallaxes) allows for detections of OC members (including their tidal tails; \citealt{Roeser2019a, Roeser2019b}) and the discovery of thousands of new clusters and moving groups almost entirely from proper-motion measurements \citep[e.g.][]{GaiaCollaboration2018, Cantat-Gaudin2018, Kounkel2019, Meingast2019}. The {\it Gaia} photometry \citep{Evans2018} allows us to characterise these objects in detail through their colour-magnitude diagrams.

In this work, we used the homogeneously derived parameters for 1\,867 {\it Gaia}-detected clusters recently published by \citetalias{Cantat-Gaudin2020b}. For that catalogue, the main cluster parameters age, distance modulus, and extinction were computed from the observed {\it Gaia} DR2 parallaxes and $G$ versus $(G_{BP}- G_{RP})$ colour-magnitude diagrams by a multi-layer-perceptron neural network trained on a set of 347 OCs with well-determined parameters (primarily from \citealt{Bossini2019}). The cluster membership lists were mostly taken from \citet{Cantat-Gaudin2020a} and \citet{Castro-Ginard2020}. The typical $\log t$ uncertainties derived by the neural network amount to 0.15-0.25 for clusters younger than 300 Myr, and 0.1-0.15 for clusters older than that. For details of the method, we refer to \citetalias{Cantat-Gaudin2020b}.

Recent studies of the CAF \citep{Joshi2016, Piskunov2018, Krumholz2019} have relied on the cluster data compiled in the latest version of the Milky Way Star Cluster survey \citep[MWSC;][]{Kharchenko2016}. A substantial fraction of objects contained in this catalogue, however, could not be confirmed with {\it Gaia} DR2 \citep{Cantat-Gaudin2018, Cantat-Gaudin2020a}; all the putatively old, high-latitude, inner-galaxy OC candidates, many dubious New General Catalogue (NGC; \citealt{Dreyer1888}) objects, and about 50\% of the old nearby FSR cluster candidates of \citet{Froebrich2007} are among the objects that could not be confirmed. In the following analysis, we thus compare our {\it Gaia} results both to the \citet{Kharchenko2013} version of the MWSC and to the latest version of that survey.

To illustrate the transformative power of {\it Gaia} DR2 on the field, Fig. \ref{fig:xymap} compares the distribution of OCs in heliocentric Cartesian coordinates derived from the MWSC catalogue with the distribution obtained from the new catalogue of {\it Gaia}-detected OCs of \citetalias{Cantat-Gaudin2020b}. For a deeper discussion of the structures emerging from this figure, we refer to the latter paper. Our main objective is to estimate the (age-dependent) completeness of the new catalogue to determine the CAF.

\begin{figure}
\begin{center} 
\includegraphics[width=.495\textwidth]{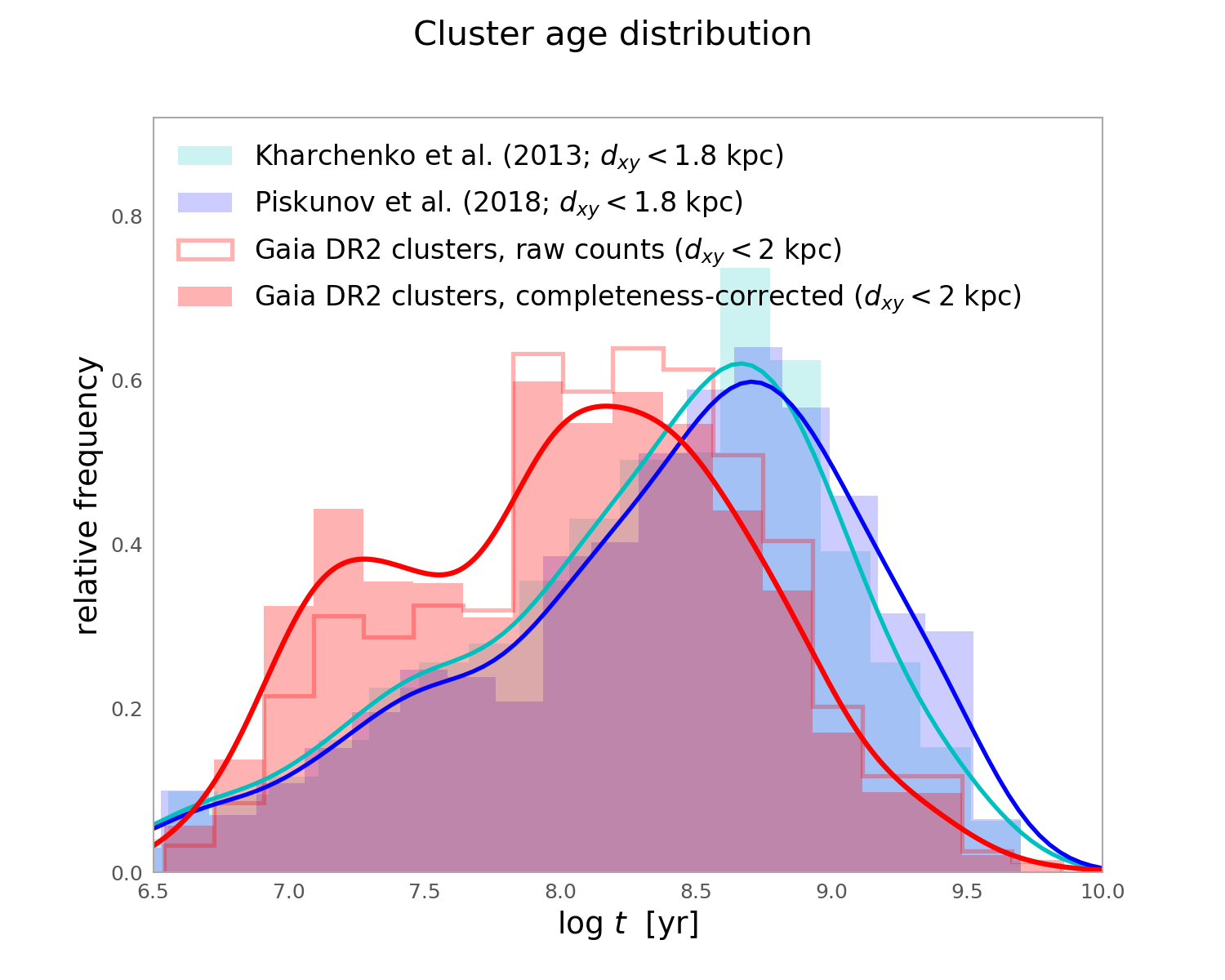}\\
\includegraphics[width=.495\textwidth]{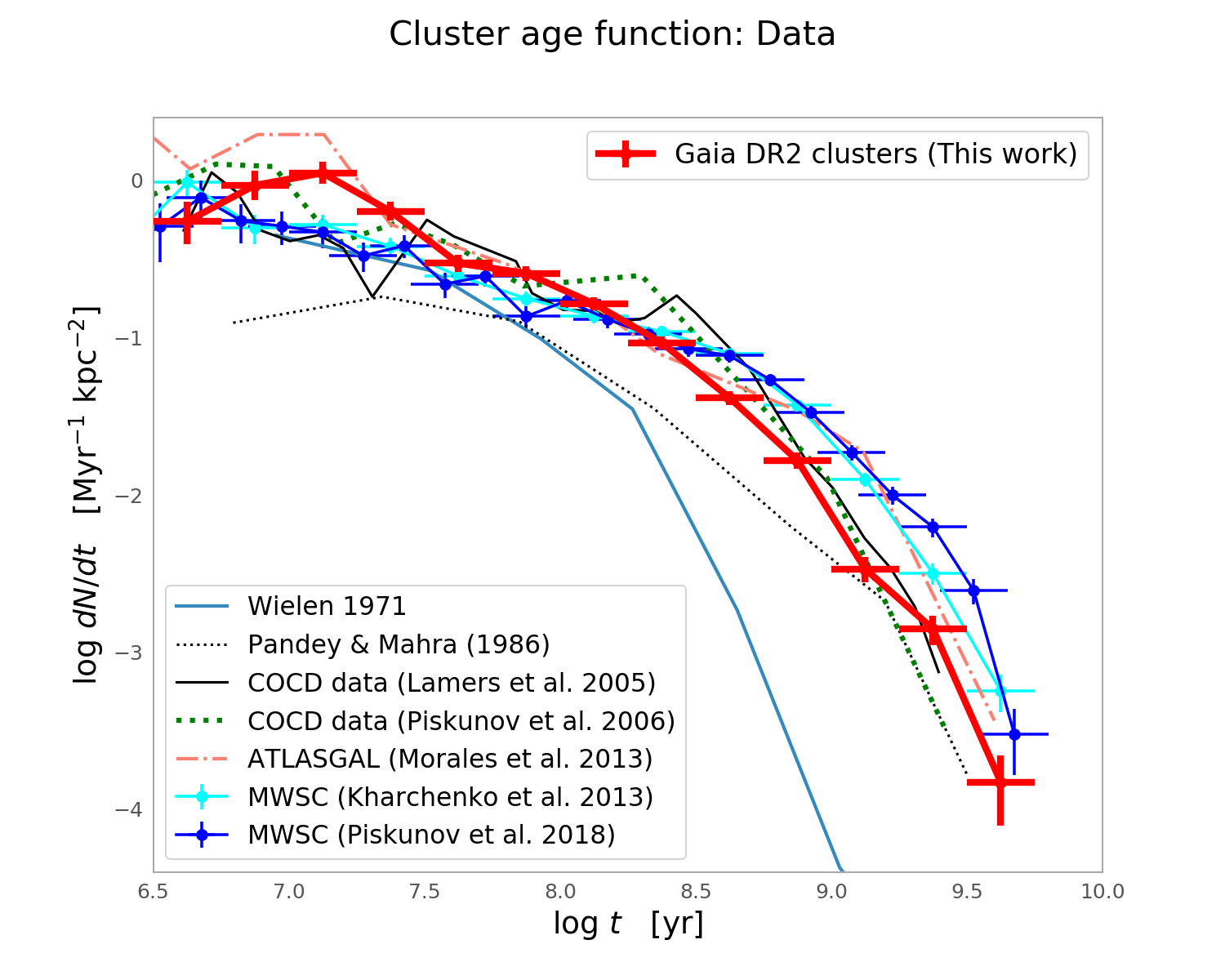}\\
\includegraphics[width=.495\textwidth]{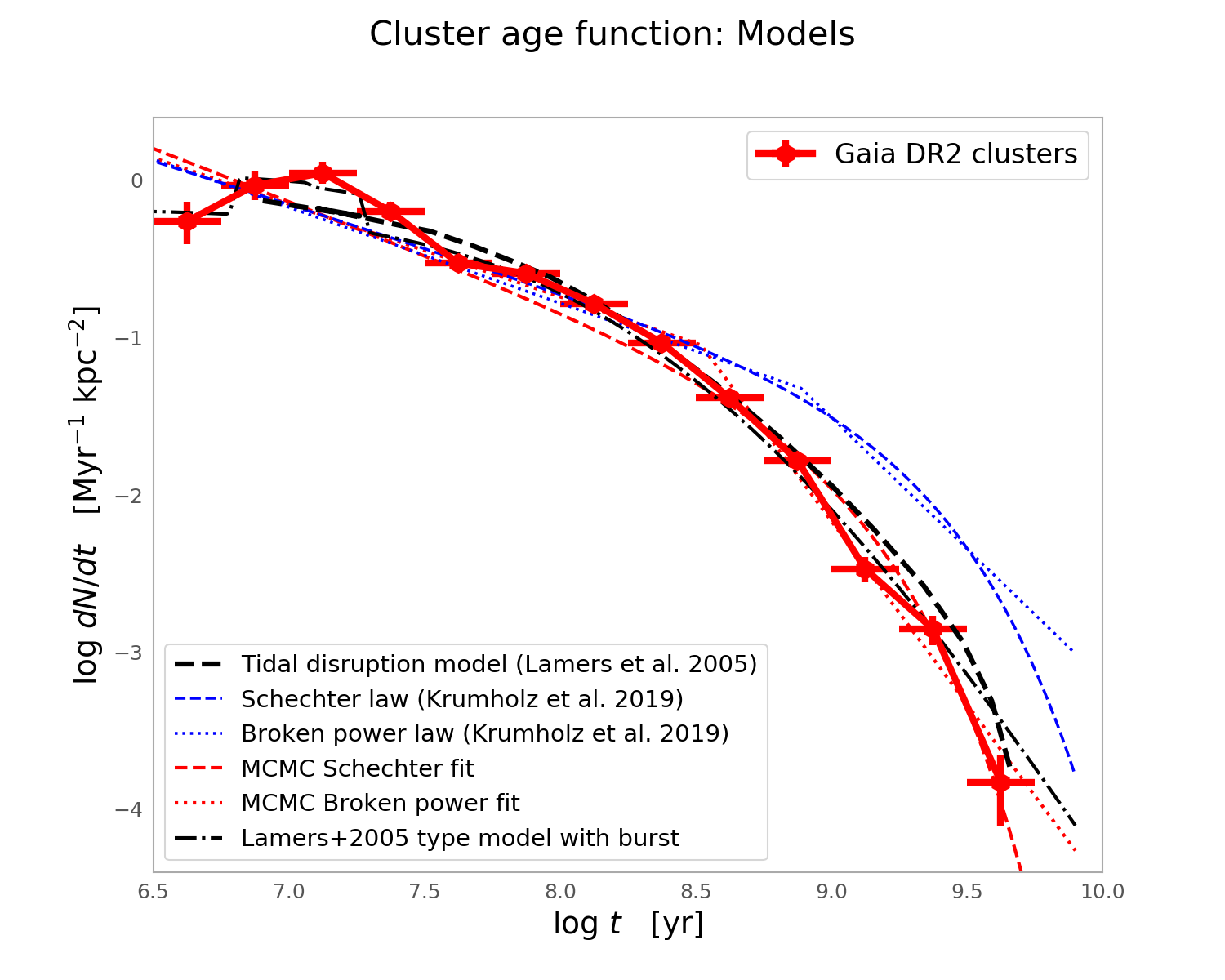}
\caption{\label{fig:agedist} 
Age distribution for Galactic open clusters in the solar vicinity. Top panel: Normalised histograms and kernel-density estimates. The cyan and blue distributions show the results from the MWSC survey (\citealt{Kharchenko2013} and \citealt{Piskunov2018}, respectively); the red distribution shows our {\it Gaia} DR2-derived results. Middle and bottom panels: Observational CAF determinations for 
the extended solar neighbourhood, from \citet{Wielen1971} to our completeness-corrected {\it Gaia} DR2-based census (red). Error bars include Poissonian uncertainties in the number of clusters per bin and systematic uncertainties from the completeness correction (see Appendix \ref{sec:appendix}). 
Bottom panel: CAF comparison to models, as indicated in the legend.
} 
\end{center}
\end{figure}

To correct for our incomplete view of the Galactic OC population, we need to quantify how selection biases affect our samples. The different aspect of the OC distributions in the right column of Fig. \ref{fig:xymap} already suggests that the present {\it Gaia} DR2 census is unlikely to be complete to a fixed limit, as was frequently assumed for the MWSC catalogue (the dashed grey circle in the left-column panels denotes the 1.8 kpc completeness limit used by \citealt{Kharchenko2016, Joshi2016}, and \citealt{Piskunov2018}).

In this work, we estimate the age-dependent completeness of the {\it Gaia} DR2 cluster census within a cylinder of radius $d_{xy}=2$ kpc (right column of Fig. \ref{fig:xymap}). The analysis can be retraced in more detail in Appendix \ref{sec:appendix}. In a nutshell, we take into account two effects. First, to account for undetected clusters, we use the OC recovery experiment performed for the latest Galactic-plane OC search of \citet{Castro-Ginard2020} to estimate the detection efficiency of their conservative method as a function of distance, sky region, and age. The second incompleteness effect stems from uncharacterised clusters: it was not possible to infer physical parameters in \citetalias{Cantat-Gaudin2020b} for all {\it Gaia}-detected OCs. Within the 2 kpc cylinder, however, this effect is minor: only 32 non-characterised clusters have Bayesian parallax distances smaller than 2 kpc. Estimating their age distribution using the values of \citet{Kharchenko2013}, we find that they are mostly younger than log $t=7.5$. The combined completeness fractions for each age bin are given in Fig. \ref{fig:xymap}.

\section{The post-{\it Gaia} DR2 cluster age function} \label{sec:caf}

Having established the completeness limits of the {\it Gaia} DR2 cluster sample, we can now determine the age distribution and the CAF. The top panel of Fig. \ref{fig:agedist} shows the histogram and a kernel density estimate of the logarithmic age distribution for the {\it Gaia} DR2 census within a 2 kpc cylinder around the Sun. For comparison, we also show the results obtained with the original MWSC catalogue \citep{Kharchenko2013} and the latest version used by \citet{Piskunov2018} and \citet{Krumholz2019}. From Fig. \ref{fig:agedist} we can already appreciate some important differences to these pre-{\it Gaia} works. The peak of the distribution lies around $\log t \sim8.2$, and although the {\it Gaia} census is much more complete for old OCs, we see a lot less of those objects. 

The typical metric for the cluster age distribution, used both by the Galactic and the extragalactic community, is the CAF, which is the number of clusters per unit of age in logarithmic age bins. Following the method of \citet{Piskunov2018}, we derived the CAF for the MWSC and the {\it Gaia} samples. Our results are shown in the middle panel of Fig. \ref{fig:agedist}. In this panel, we also show some of the literature results compiled by \citet{Piskunov2018}: namely \citet{Wielen1971, Pandey1986, Lamers2005, Piskunov2006}, and \citet{Morales2013}.

In the bottom panel of Fig. \ref{fig:agedist}, we compare our data to several models. In particular, these are two of the cluster destruction models presented by \citet{Lamers2005}, a fit to the \citet{Lamers2006} model, and the results of our fits to two simple analytical functions; these are all performed with the Markov chain Monte Carlo sampler {\tt emcee} \citep{Foreman-Mackey2013}. 

We confirm the conclusion of \citet{Krumholz2019} that the Milky Way CAF is closely fitted by a Schechter function or a broken power law. The fit parameters for those functions, however, have to be revised. In particular, we obtain best-fit values of 
$\alpha_T = -0.65^{+0.10}_{-0.10}, \log t^{\ast}= 9.30^{+0.07}_{-0.06}$ for the Schechter case (\citealt{Krumholz2019}: $\alpha_T = -0.55, \log t^{\ast}=9.59$), and $\alpha_1 = -0.56^{+0.16}_{-0.11}, \alpha_2 = -2.34^{+0.29}_{-0.36}, \log t^{\rm break}=8.49^{+0.21}_{-0.21}$ for the case of a broken power law (\citealt{Krumholz2019}: $\alpha_1 = -0.61, \alpha_2 = -1.67, \log t^{\rm break}=8.89$). 
Our basic conclusion is that the downturn in the CAF occurs at lower ages (by a factor $2-3$) and the slope beyond the break is steeper.

\section{Discussion} \label{sec:discussion}

The CAF is the marginalised probability distribution of the full Galactic OC distribution function. Until the first mass estimates for {\it Gaia} clusters become available, it is our best tool to study the physics of OC formation and destruction. The new homogeneous age catalogue of {\it Gaia} DR2-detected OCs of \citetalias{Cantat-Gaudin2020b} allows us to probe this observable with better precision and accuracy than ever before. 

Our measurements, summarised in Fig. \ref{fig:agedist}, rule out the old-age-heavy CAF obtained in recent years from the MWSC catalogue, which contains a significant number of allegedly old false positives. Instead, our new CAF determination is in line with earlier measurements (e.g. \citealt{Lamers2005, Piskunov2006, Morales2013}; for a detailed discussion of the history of Milky Way CAF measurements, we refer to Sect 3.3 of \citealt{Piskunov2018}).

This also implies that some cluster formation and destruction models from the pre-MWSC era are still compatible with our measurements. In particular, this is the case for the model of \citet{Lamers2005} with a typical destruction timescale for a $10^4{\rm M}_{\odot}$ OC of $t_4\sim 1.5$ Gyr, which we show in Fig. \ref{fig:agedist}. Those authors modelled the cluster destruction as $t_{\rm dis} \propto t_4\cdot (M_{\rm ini}/10^4{\rm M}_{\odot})^{\gamma}$, with $\gamma\approx0.62$ and a star formation rate in clusters of $\sim500$ M$_{\odot}$ Myr$^{-1}$kpc$^{-2}$. Surprisingly, almost all of our CAF data points (except for the lowest age bin) are consistent with the \citet{Lamers2005} model within $1\sigma$.
In addition, in accordance with \citet{Morales2013}, we find a hint of a short bump in the cluster formation rate at very young ages, around $6-20$ Myr (dash-dotted curve in Fig. \ref{fig:agedist}). The proximity of our data to the CAF obtained by \citet{Morales2013} from ATLASGAL radio data \citep{Schuller2009} of mostly embedded clusters towards the inner Galaxy also suggests little change in the cluster destruction rate within a few kiloparsec from the Sun.

\citet{Lamers2006} parametrised the destruction time of initially bound OCs in the solar neighbourhood, taking into account four processes in the life of OCs: stellar evolution, tidal disruption by the Galactic gravitational field, shocking by spiral arms, and (most importantly) encounters with giant molecular clouds. These authors showed that the observed CAF depends on the destruction timescale, the cluster formation rate, and the cluster initial mass function. In the absence of cluster masses, however, we find that these parameters are still degenerate and thus refrain from reporting fit values for such a model. The current cluster formation rate can in principle be read off the lowest age bin ($0.55^{+0.19}_{-0.15}$ Myr$^{-1}$kpc$^{-2}$); this value, however, is most affected by our completeness corrections and should be treated cautiously. 

To convert the cluster formation rate to the star formation rate in clusters, we assume a mean initial mass of $\sim 300-600\,{\rm M}_{\odot}$; the exact value depends on the cluster initial mass function and the lower-mass limit of our sample. Assuming this mean initial mass, we obtain a star formation rate in clusters of $250^{+190}_{-130}$ M$_{\odot}$ Myr$^{-1}$kpc$^{-2}$, which, when compared to the total star formation rate in the solar vicinity ($1600^{+700}_{-400}$ M$_{\odot}$ Myr$^{-1}$kpc$^{-2}$; \citealt{Mor2019}), suggests that only $16^{+11}_{-8}$\% of the stars in the solar vicinity form in bound clusters (see also \citealt{Adamo2020, Ward2020}). 

\citet[][Sect. 2.3]{Krumholz2019} reviewed determinations of the CAF for external galaxies (based on unresolved cluster observations) and compared these determinations to the local CAF obtained from the \citet{Piskunov2018} data. These authors note that the typical power-law index $\alpha_T$ for the Milky Way seemed to steepen quickly around ages of $\sim10^9$ Gyr, in stark contrast to other galaxies (their Fig. 6). Our revised OC census based on {\it Gaia} DR2 seems to bring the Milky Way back in line with most other spiral galaxies, including M31, for the age range $7.5<\log t<9.5$.

The comparison to extragalactic samples, however, is still biased in two other ways \citep{Krumholz2019}: First, our Milky Way sample is still limited to $\sim 2-3$ kpc from the Sun and consists entirely of low-mass clusters ($M\lesssim10^3 {\rm M}_{\odot}$), while extragalactic samples are dominated by the most massive clusters (usually $M > 10^{3.5} {\rm M}_{\odot}$). Second, ages for extragalactic clusters are derived from integrated photometry, whereas ours have been derived from high-precision {\it Gaia} DR2 colour-magnitude diagrams and parallaxes, thus making our measurement a new benchmark for extragalactic studies as well.

We look forward to the next {\it Gaia} data releases, which will enable an even deeper characterisation of thousands of Galactic open clusters, eventually also allowing for precise determinations of cluster masses. The joint mass and age distribution of the Milky Way, as well as variations in the CAF as a function of position in the Galaxy, will then allow us to test the limits of the most updated cluster formation and destruction models. 

\bibliographystyle{aa} 
\bibliography{clusteragefunction_letter_final}

\begin{acknowledgements}
We thank Mercè Romero-Gómez and Cesca Figueras for their comments. 
This work has made use of data from the European Space Agency (ESA) mission \textit{Gaia} (www.cosmos.esa.int/gaia), processed by the \textit{Gaia} Data Processing and Analysis Consortium (DPAC, www.cosmos.esa.int/web/gaia/dpac/consortium). Funding for the DPAC has been provided by national institutions, in particular the institutions participating in the \textit{Gaia} Multilateral Agreement. This work was partially supported by the Spanish Ministry of Science, Innovation and University (MICIU/FEDER, UE) through grant RTI2018-095076-B-C21, and the Institute of Cosmos Sciences University of Barcelona (ICCUB, Unidad de Excelencia ’Mar\'{\i}a de Maeztu’) through grant CEX2019-000918-M. FA is grateful for funding from the European Union's Horizon 2020 research and innovation programme under the Marie Sk\l{}odowska-Curie grant agreement No. 800502.

The preparation of this work has made use of TOPCAT \citep{Taylor2005}, NASA's Astrophysics Data System Bibliographic Services, as well as the open-source Python packages \texttt{Astropy} \citep{Astropy2018}, \texttt{NumPy} \citep{VanderWalt2011}, \texttt{scikit-learn} \citep{Virtanen2019}, and \texttt{emcee} \citep{Foreman-Mackey2013}. The figures in this paper were produced with Matplotlib \citep{Hunter2007}. 
\end{acknowledgements}

\appendix
\section{Completeness of the OC census} \label{sec:appendix}

In this appendix, we characterise the completeness of the OC catalogue of \citet{Cantat-Gaudin2020b}. 

The probability of detecting a resolved Galactic star cluster depends on fundamental cluster parameters, such as its total mass, age, distance, or extinction. However, we may also invoke an additional dependence on other, slightly more subtle parameters, such as the contrast in the proper-motion diagram, the number of bright stars, and the amount of differential extinction. We are facing the additional challenge that the true underlying spatial distribution of OCs is complex and unknown, and that the transition between bound clusters, associations, and dissolving structures becomes increasingly unsharp (e.g. \citealt{Kounkel2020, Ward2020}).

The Gaia DR2-based catalogue of \citetalias{Cantat-Gaudin2020b} is a clean OC catalogue, limited by the number of stars above a certain magnitude threshold ($G < 17...18$, depending on the provenance of the cluster detection). In contrast to the case of unresolved extragalactic clusters, the number of stars above the magnitude detection limit only decreases very little with mass as long as the turn-off is above the detection limit (see e.g. Fig. 2 of \citealt{Gieles2007}). We therefore opt to parametrise the selection function of our OC catalogue as a function depending primarily on Galactic longitude $l$, latitude $b$, planar distance $d_{xy}$, log $t$, and extinction $A_V$. 

We make the following approximation:
\begin{align*}
&p({\it Gaia}\, {\rm DR2\ cluster\, has\, a\, CGa20\, age})\\
&= p({\rm ANN\ converged} | {\rm cluster\, detection}) \cdot p({\rm cluster\, detection}) \\
&\approx p({\rm ANN\ converged} | d_{xy}, \log t)  \cdot p({\rm cluster\, detection} | l, b, d_{xy}, \log t)\\
\end{align*} 
In other words, the completeness of the age census of {\it Gaia} DR2 OCs depends on our ability to derive OC parameters for detected clusters and our detection efficiency. 

\begin{figure}
\begin{center}
\includegraphics[width=.495\textwidth]{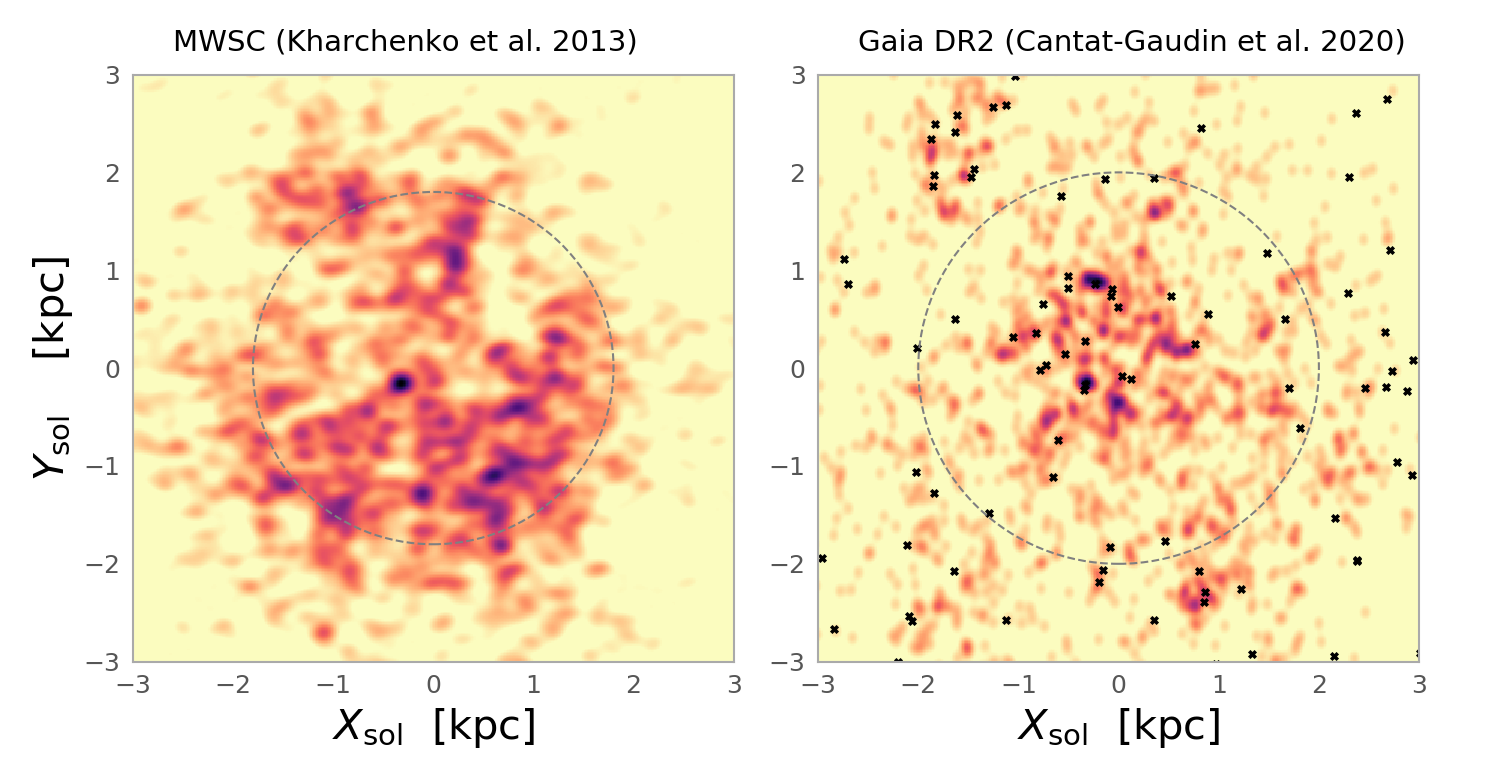}
\caption{\label{fig:tristan}
Distribution of Galactic OCs using the MWSC (left) and the {\it Gaia} DR2-derived catalogue (right), respectively. The crosses in the right panel indicate the positions of the clusters for which no physical parameters could be obtained in \citet{Cantat-Gaudin2020b}.
} 
\end{center}
\end{figure}

For the first factor (completeness of the ANN-derived parameters), we simply determine the fraction of clusters within $d_{xy}=2$ kpc for which no physical parameters were obtained in \citet[][see Fig. \ref{fig:tristan}]{Cantat-Gaudin2020b}. This fraction is very low ($\lesssim 1\%$) for $\log t>7$ and increases to 23 \% for the youngest age bin (using the age estimates of \citealt{Kharchenko2013}).

\begin{table}
\caption{Completeness estimates as a function of age for the \citetalias{Cantat-Gaudin2020b} catalogue out to $d^{\rm lim}_{xy} = 2$ kpc.}
\label{appendixtable}
\begin{tabular}{l|c|c|c}
log t & ANN          & Detection              & Combined               \\
$[$yr$]$  & completeness & completeness           & completeness     \\
\hline
6.5   & 0.77         & $0.52^{+0.08}_{-0.07}$ & $0.40^{+0.08}_{-0.07}$ \\
7     & 0.88         & $0.54^{+0.08}_{-0.07}$ & $0.48^{+0.08}_{-0.07}$ \\
7.5   & 0.99         & $0.65^{+0.05}_{-0.05}$ & $0.64^{+0.05}_{-0.05}$ \\
8     & 0.99         & $0.76^{+0.05}_{-0.04}$ & $0.75^{+0.05}_{-0.04}$ \\
8.5   & 1            & $0.81^{+0.04}_{-0.04}$ & $0.81^{+0.04}_{-0.04}$ \\
9     & 1            & $0.85^{+0.05}_{-0.04}$ & $0.85^{+0.05}_{-0.04}$ \\
9.5   & 1            & $0.88^{+0.05}_{-0.06}$ & $0.88^{+0.05}_{-0.06}$ \\
10    & 1            & $0.88^{+0.05}_{-0.06}$ & $0.88^{+0.05}_{-0.06}$ \\
\end{tabular}
\end{table}

\begin{figure*}
\begin{center} 
\includegraphics[width=.8\textwidth]{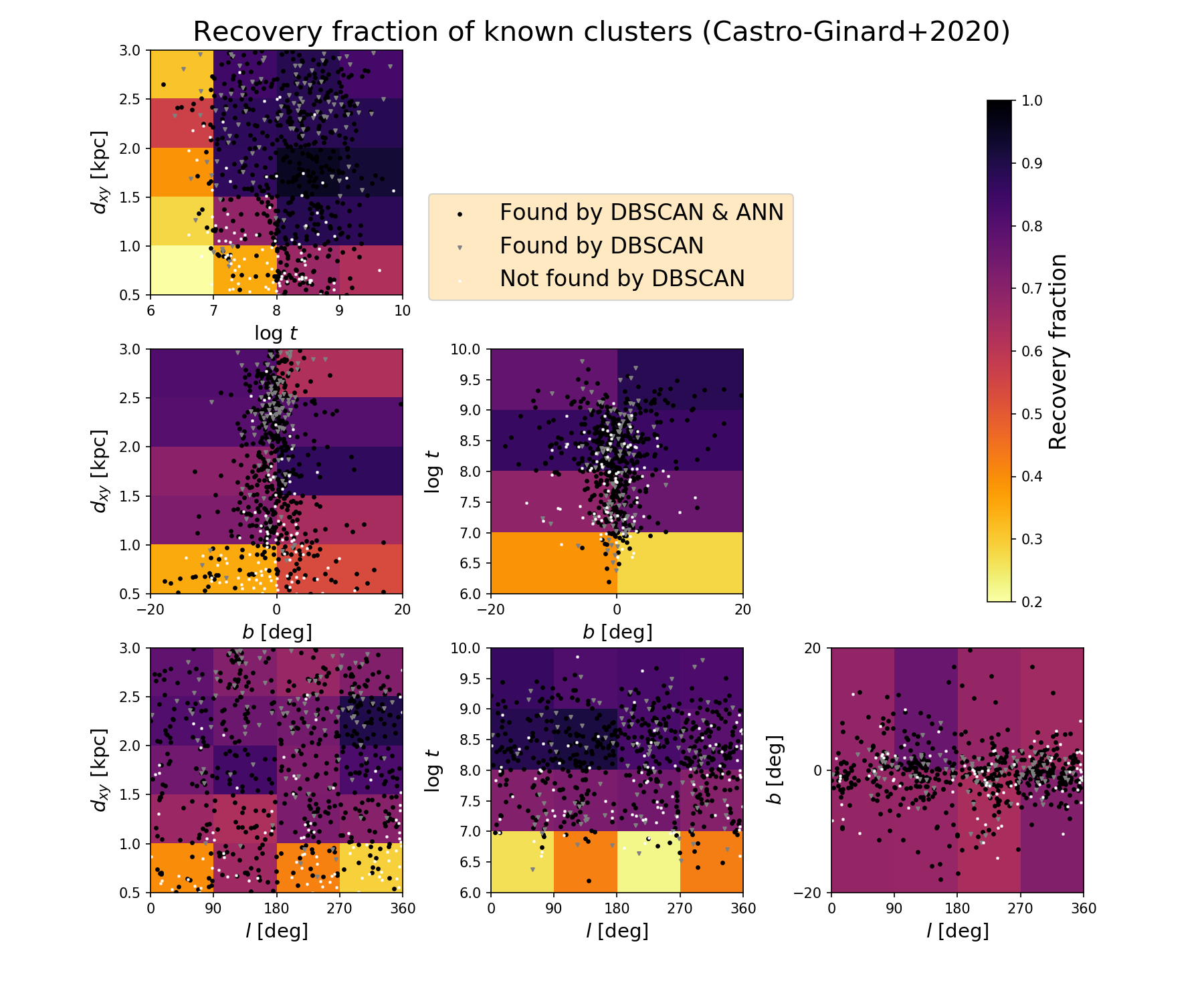}
\caption{\label{fig:alfred}
Cluster recovery fraction of the Galactic plane search of \citet{Castro-Ginard2020}, as a function of Galactic longitude, latitude, planar distance, and age. In each panel, the black, grey, and white symbols denote \citet{Cantat-Gaudin2018} clusters that were recovered, half-recovered, or not recovered, respectively, by the blind search of \citet{Castro-Ginard2020}.
} 
\end{center}
\end{figure*}

To estimate the second factor, we use the experiment carried out by \citet{Castro-Ginard2020}. These authors investigated the recovery fraction of their catalogue by comparing the blind-search detections to the OC list of \citet{Cantat-Gaudin2018}. They categorised these clusters into recovered, half-recovered, and non-recovered; see [Fig. 1 of \citealt{Castro-Ginard2020}]. In this work, we use these weights to determine the OC recovery fraction in wide bins of [$l, b, d_{xy}, \log t$].
We find that the recovery fraction does not depend significantly on extinction (beyond the intrinsic correlation of this parameter with distance). The results are shown in Fig. \ref{fig:alfred}. Since the blind search of \citet{Castro-Ginard2020} was not optimised for nearby clusters, we focus on the distance range $d_{xy}>1$ kpc.

Now we can define a completeness fraction as a function of $\log t$ out to a limiting distance $d_{xy}^{\rm max}$ by numerically integrating the selection fraction over the $d_{xy}, l$ and $b$ dimensions, and linear interpolation between the $\log t$ bins. We also estimate the uncertainty associated with the completeness estimates. To do this, we can assume Poissonian distributions of the counts in each age bin (for each group: non-detected, half-detected, and detected). This allows us to determine the completeness uncertainties numerically.

The implicit assumption behind our approach is that the experiment of \citet{Castro-Ginard2020} is representative for the full \citetalias{Cantat-Gaudin2020b} catalogue. This is debatable, but the best we can do at present. On the one hand, this estimate is pessimistic because we only estimate the completeness of one method (DBSCAN), while the full catalogue of Gaia-detected OCs was compiled from various search methods (e.g. traditional OCs, serendipitous discoveries, and blind machine-learning searches.). It is also pessimistic in the sense that the blind search of \citet{Castro-Ginard2020} has added more than 500 new OCs. On the other hand, we may argue that the DBSCAN completeness test of \citet{Castro-Ginard2020} may have been slightly optimistic since the sample of \citet{Cantat-Gaudin2018} may not be fully representative of the underlying population.

The results of our completeness calculations are given in Table \ref{appendixtable}. The second column contains the first factor (physical parameter completeness), the third column contains the second factor (cluster recovery fraction), and the last column the product of the two. The associated systematic uncertainties have been propagated into the uncertainties of the CAF (e.g. the error bars in Fig. \ref{fig:agedist}) and all derived quantities.

\end{document}